\documentclass[]{interact}

\usepackage{amssymb,amsmath,amsfonts,eurosym,geometry,ulem,graphicx,caption,color,setspace,comment,footmisc,caption,pdflscape,array,hyperref}

\usepackage{epstopdf}

\usepackage{natbib}
\usepackage{multirow}
\usepackage{subcaption}

\theoremstyle{plain}

\theoremstyle{definition}

\theoremstyle{remark}

\DeclareMathOperator{\Var}{Var}
\DeclareMathOperator{\Cov}{Cov}

\begin{document}


\title{Weighted model calibration with spatial conditional information}

\author{
\name{Michele Nguyen \textsuperscript{a}\thanks{Corresponding author: Michele Nguyen (michele.nguyen@ntu.edu.sg)}, Maricar Rabonza \textsuperscript{b} and David Lallemant \textsuperscript{b}}
\affil{\textsuperscript{a}Lee Kong Chian School of Medicine and School of Physical \& Mathematical Sciences, Nanyang Technological University, Singapore; \textsuperscript{b} Asian School of the Environment, Nanyang Technological University, Singapore}
}

\maketitle

\begin{abstract}
Cost functions such as mean square error are often used in environmental model calibration. These treat observations as independent and equally important even though model residuals exhibit spatial dependence and additional observations near existing points do not provide as much information on the system as those elsewhere. To address this issue, we develop a method to derive calibration weights based on spatial conditional information. Using simulation experiments with Gaussian processes and the Tephra2 volcanic tephra dispersion model, we show that the additional accuracy and precision from weighted inference increases with the degree of observation clustering and spatial dependence present. To demonstrate real-world relevance, the methods are applied to tephra load observations from the 2014 eruption of the Kelud volcano in Indonesia.     
\end{abstract}


\section{Introduction}

Cost functions and goodness-of-fit metrics quantify the level of agreement between observations and modelled quantities. These are commonly used in the environmental sciences to calibrate and evaluate models. In the context of urban cooling, for example, \cite{Zawadzka2021} calibrated the InVEST model by maximising the coefficient of determination (R$^2$) between modelled and observed land surface temperatures. To model streamflow and sediment yields, \cite{Wang2014} calibrated the Agricultural Policy Environmental eXtender (APEX) model using an objective function combining the Nash-Sutcliffe model efficiency coefficient (NSE) and percent bias. 
\\
Despite the wide variety of cost functions and their ubiquitous use in environmental modelling, majority share a much overlooked but non-trivial assumption: the statistical independence of observed data. This means that after accounting for a deterministic trend via a process-based model or secondary variables, the remaining model residuals are seen as independent samples from some common error distribution. This assumption is implicit in the expression of the cost functions as a sum of error contributions from individual observations. \cite{Rabonza2022} show that common cost functions such as mean square error (MSE), chi square error and mean square logarithmic error (MSLE) are proportional to the negative log-likelihoods of independent Gaussian distributions. In the same way, mean absolute error and mean absolute percentage error are related to independent Laplace distributions.
\\
Counter to this independence assumption, correlation is often observed in model residuals. In line with Tobler's first law of geography \citep{Tobler1970}, model residuals from observations near to each other tend to be more similar than those further away. While proximity can be defined via time, space or any other measurement characteristic, we focus on proximity in space. Spatial correlation can arise due to an inherent property of the variable (for example, in physical dispersal and growth), external spatially structured environmental factors or the study design \citep{Haining2001, Kissling2008, Gaspard2019}. Spatial correlation in model residuals indicates that the spatial characteristics in the data are not well explained by the fitted model and is a sign of model inadequacy (see for example, \cite{Valcu2010}).  
\\
When spatial correlation is not accounted for in model calibration (as is often the case), there are implications on the validity of probabilistic results and the statistical tests which are used to select and/or evaluate the model. Random sampling in the presence of dependence leads to a subtle form of subsampling since observations within a certain distance tend to be similar and we are unlikely to sample the full range of possible values \citep{Dale2014}. By erroneously assuming independence and our ability to cover the whole sample space, we assume that we have more information about the population than we actually do. This inflates our effective sample size for the given dataset, leading to underestimated standard errors of our derived estimates, inflated test statistics for hypothesis tests and increased chances of Type I error, i.e. false positives. Hence, failure to account for spatial correlation within a dataset can lead to suboptimal variable selection and model misspecification \citep{Bahn2006, Dormann2007}.
\\
The effect of spatial correlation on variable selection and parameter estimation has been seen in the ecological literature. Through the reanalysis of ecological datasets, \cite{Dormann2007} observed an average of 25\% shifts in the estimated regression coefficients when they move from a non-spatial to a spatial model. However, this does not mean that regression coefficients from non-spatial models are biased. While neglecting spatial correlation leads to underestimated standard errors of our estimates for a given dataset, repeating the estimation over multiple datasets can give estimates which on average are equal to the true coefficient value. \cite{Beale2007} illustrate this for the case where the true model structure is an additive spatial error model (i.e. a linear model plus spatial error). Mathematically, it can be shown that both ordinary least squares and spatial generalised least squares (GLS) give unbiased regression coefficients. Instead of unbiasedness, the advantage of modelling spatially via GLS was the higher precision of parameter estimates in the presence of spatial correlation. This was observed through simulation experiments and is dependent on the type of spatial correlation present as well as the assumed spatial model \citep{Kissling2008}. 
\\
Many spatial models have been proposed to explicitly model spatial dependence. Examples include simultaneous autoregressive (SAR) models which encode spatial dependence through lags of the response and/or errors \citep{Kuhn2007}, spatial eigenvector filtering \citep{Kim2016}, autocovariate models \citep{Dormann2007} and generalised least squares (GLS) or spatial generalised linear mixed models (GLMM), where space treated as continuous instead of discrete. These models, however, do not share the same advantages of process-based models in terms of being directly related to the physics of the environmental phenomena. 
\\
In this paper, we propose a weighted calibration approach using spatial conditional information to account for spatial dependence in process-based models. Accounting for spatial correlation has been shown to be important for both model prediction and validation. For example, \cite{Rabonza2022} harnessed residual spatial correlation to generate more localised volcanic tephra estimates in line with observations. Unlike those obtained from the fitted process-based model which generally show long-range trends, these are more useful for damage assessments. In the context of model validation, multiple studies show that predictive performance can be overinflated when we do not accounting for spatial dependence \cite{Karasiak2022, Kattenborn2022, Ploton2020}. By developing an approach to account for spatial correlation in model calibration, we contribute to the consideration of spatial correlation in every modelling step. 
\\
In Section \ref{sec:theory}, we present the weighted calibration approach. This involves weighted cost functions and the derivation of the calibration weights using spatial conditional information and variograms. In Section \ref{sec:GPsim}, we conduct simulation experiments using Gaussian processes to generate spatial dependence. Similar to \cite{Beale2007}, we examine the benefits of accounting for spatial dependence through weighting in terms of bias and precision of parameter estimates. While \cite{Beale2007} focused on raster data and estimated the slope of a linear model in the presence of spatial correlation, we focus on point observations which are more typical of field surveys and estimate the mean of the Gaussian process. This is because process-based models will be used to determine deterministic long-range trends in place of the covariates in a linear model. By comparing estimates from weighted and unweighted inference as well as from a full spatial likelihood optimisation, we investigate the importance of weighting in the presence of observation clustering as well as different levels of spatial dependence. With these insights, we apply our methods to the calibration of a process-based model, the Tephra2 model for volcanic tephra dispersion in Section \ref{sec:Tephra2}. Since the calibration weights depend on the fitted model, this involves iteratively reweighting the observations during the Tephra2 calibration process.  We show that weighted calibration leads to significantly different parameter estimates and modelled tephra load surfaces for the 2014 eruption of the Kelud volcano. Through simulation experiments, we also show that the benefits of weighting in terms of bias and precision persist despite the strong interactions and correlations between the Tephra2 model parameters. In Section \ref{sec:discussion}, we discuss our results and their implications for more general environmental modelling. Finally, we point towards future directions for research. 

\clearpage

\section{Weighted model calibration} \label{sec:theory}

\subsection{Weighted cost functions}

Cost functions which are often used for model calibration can be expressed as:

\begin{equation}
    C(\mathbf{y}, \hat{\mathbf{y}}) = \sum\limits_{i = 1}^{n} g(\hat{y}_{i} - y_{i}),
\end{equation}

where $\mathbf{y} = (y_{i}, \dots, y_{n})$ is the vector of $n$ observations and $\hat{\mathbf{y}} = (\hat{y}_{i}, \dots, \hat{y}_{n})$ denotes their corresponding model estimates. The function $g$ describes the way the individual model residuals (the difference between the modelled and observed values) contribute to the cost function. In this formulation, since the cost function is a simple sum of these contributions from the $n$ observations, the observations are thought to provide equal and independent information. This assumption is implicit for mean square error (MSE) where $g(\hat{y}_{i} - y_{i}) = \frac{1}{n}(\hat{y}_{i} - y_{i})^2$ and mean absolute error where $g(\hat{y}_{i} - y_{i}) = \frac{1}{n}|\hat{y}_{i} - y_{i}|$.

In the case of MSE, the cost function is directly proportional to the negative log-likelihood of independent Gaussian random variables (less the scalar constant $n\log(\sigma\sqrt{2\pi})$):

\begin{equation}
    MSE = \frac{1}{n}\sum\limits_{i = 1}^{n} (\hat{y}_{i} - y_{i})^2 \propto \frac{1}{2}\sum\limits_{i = 1}^{n} \left(\frac{\hat{y}_{i} - y_{i}}{\sigma}\right)^{2}.
\end{equation}

Hence, minimising MSE is akin to maximising the likelihood. Here, the random variables are $Y_{i} \sim N(\hat{y}_{i}, \sigma^2)$ for some standard deviation $\sigma>0$.
\\
Instead of independent Gaussian likelihoods, a full spatial likelihood can be used to express the spatial dependence in the model residuals and maximum likelihood estimation through for example, kriging can be used to estimate model parameters \citep{Chiles1999}. The likelihood, however, now takes the form of a multivariate Gaussian density involving the inverse and determinant of a $n\times n$ spatial covariance matrix. For large datasets, exact analytical methods become infeasible, especially when we need to estimate multiple parameters as in the case of process-based models.
\\
One way to approximate a dependent likelihood is to weight independent likelihoods according to some measure of their dependence \citep{Varin2011}. A weighted cost function can be expressed as:

\begin{equation}
    C_{W}(\mathbf{y}, \hat{\mathbf{y}}) = \sum\limits_{i = 1}^{n} w_{i}g(\hat{y}_{i} - y_{i}),
\end{equation}

where $w_{i}$ denotes a calibration weight related to spatial dependence. For example, weighted MSE can be defined as:

\begin{equation}
    WMSE = \frac{1}{n}\sum\limits_{i = 1}^{n} w_{i}(\hat{y}_{i} - y_{i})^2.
\end{equation}

\subsection{Quantifying spatial dependence with variograms}

In kriging, the variogram is often used to describe spatial dependence. Let $e_{i} = \hat{y}_{i} - y_{i}$, the model residual corresponding to observation $i$ at site $s_{i}$. The variogram is defined as half the expected square difference between model residuals at different locations:

\begin{equation}
    \gamma(s_{1}, s_{2}) = \frac{1}{2}\mathbb{E}\left[\left(e_{1} - e_{2}\right)^{2}\right].
\end{equation}

There is a one-to-one relationship between the variogram and the spatial covariance:

\begin{equation}
    \gamma(s_{1}, s_{2})  = \Var(e) - \Cov\left(e_{1}, e_{2}\right).
\end{equation}

\begin{figure}[tbp]
    \centering
        \includegraphics[width = \textwidth, trim = 1cm -1cm 1cm 1cm]{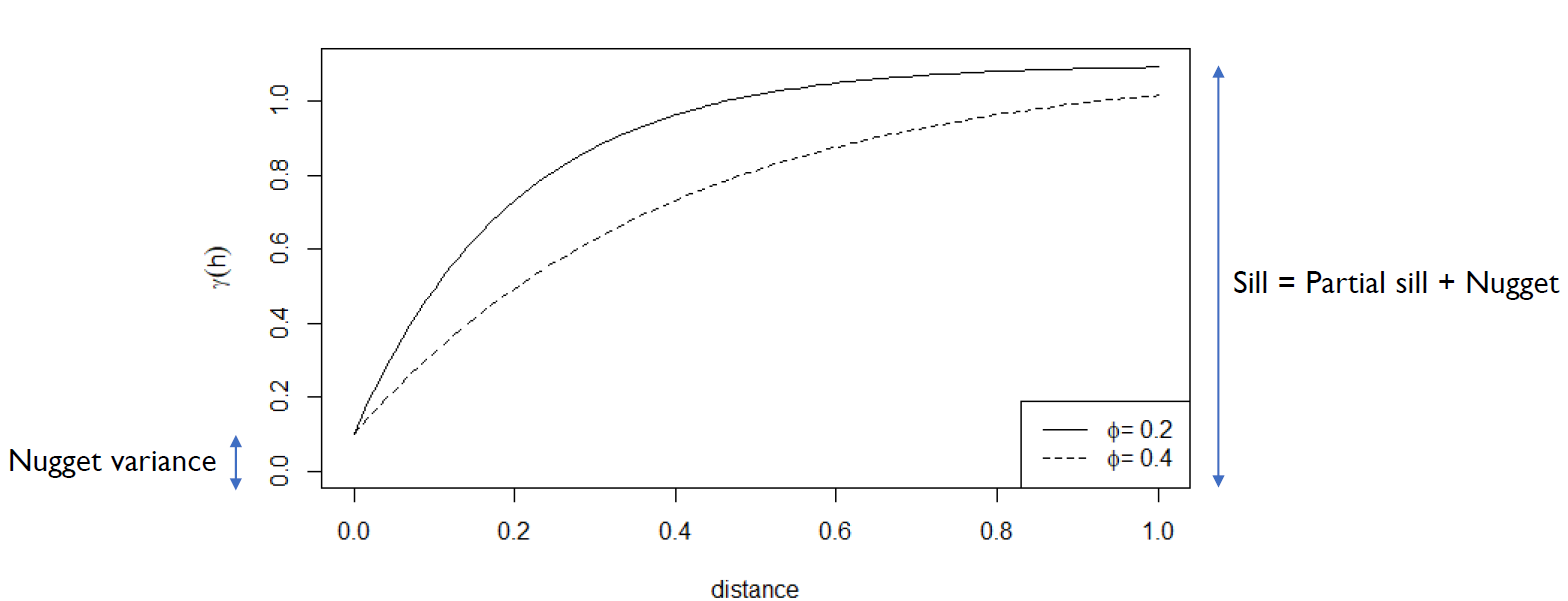}
\caption{Exponential variogram with different range parameters.}
\label{fig:vgm}
\end{figure}

Under the common assumptions of stationarity and isotropy, the variogram becomes a function of the distance between observations, $\gamma(s_{1}, s_{2}) = \gamma(h)$ where $h = |s_{1}-s_{2}|$. An example is the exponential variogram:

\begin{equation}
    \gamma(h) = \sigma^{2}\left[1 - \exp\left(-\frac{h}{\phi}\right)\right] + \tau^{2},
\end{equation}

where $\sigma^{2}$ denotes the partial sill (the amount of spatially correlated variance), $\phi$ is a range parameter (controlling the rate of correlation decay with distance) and $\tau^2$ is the nugget variance (the amount of spatially independent variance or measurement error). Figure \ref{fig:vgm} shows two exponential variograms with the same nugget variance and partial sill but different range parameters. The larger range parameter leads to a slower decay in spatial correlation.

In our analyses, we will use Mat\'ern variograms which commonly used in Spatial Statistics and which extend the exponential variogram towards greater flexibility in terms of the smoothness of the generated processes:

\begin{equation}
    \gamma(h) = \sigma^{2}\left[1 - \frac{1}{2^{\kappa-1}\Gamma(\kappa)}\left(\frac{h}{\phi}\right)^{\kappa}K_{\kappa}\left(\frac{h}{\phi}\right)\right] + \tau^{2},
\end{equation}

where $\Gamma$ denotes the Gamma function, $\kappa$ is a smoothness parameter and $K_{\kappa}$ is the modified Bessel function of the second kind.

\subsection{Weights based on spatial conditional information} 

Next, we introduce the concept of spatial conditional information (CI). This is related to the full spatial likelihood which involves spatial covariance matrices related to variograms. Based on transformations of CIs which are chosen to obtain desirable properties, we develop a weighting scheme for model calibration. Inference results from the weighted calibration will be compared with unweighted calibration in later Sections.

\subsubsection{Spatial conditional information}

When there is spatial dependence, clustered observations provide repeated information on the process of interest because their values are likely to be very similar. To quantify the information gained or the amount of uncertainty reduced by sampling a process at an additional location, we can compute its CI. As defined in Equation (18) of \cite{Wellmann2013}, the CI of a process $Y$ at its $k+1$th location is defined as: 

\begin{equation}
    H(Y_{k+1}|Y_{1},Y_{2}, \dots, Y_{k}) \\ = H(Y_{1}, Y_{2}, \dots, Y_{k}, Y_{k+1}) - H(Y_{1}, Y_{2}, \dots, Y_{k}).
\end{equation}

Here, $Y_{i}$ denotes the random variable for the process at location $i$,

\begin{equation}
H(Y_{1}, Y_{2}, \dots, Y_{k}) = -\sum p_{Y_{1}, Y_{2}, \dots, Y_{k}} \log_{2} p_{Y_{1}, Y_{2}, \dots, Y_{k}}  
\end{equation}

denotes the joint entropy for $Y_{1}, Y_{2}, \dots, Y_{k}$ and $p_{Y_{1}, Y_{2}, \dots, Y_{k}}$ is their joint distribution. 

When we use the multivariate normal distribution to model $Y_{1}, Y_{2}, \dots, Y_{k}$ as is done in a spatially dependent Gaussian process, we can calculate the joint entropy by\citep{Ahmed1989}:

\begin{equation}
H(Y_{1}, Y_{2}, \dots, Y_{k}) = \frac{1}{2}\log(|(2\pi e)\Sigma_{k}|)\\ = \frac{k}{2} + \frac{k}{2}\log(2\pi) + \frac{1}{2}\log(|\Sigma_{k}|)
\end{equation}

where $\Sigma_{k}$ is the $k\times k$ spatial covariance matrix. So, the CI of $Y_{k+1}$ given $Y_{1}, Y_{2}, \dots, Y_{k}$ is:

\begin{align}
&H(Y_{k+1}|Y_{1}, Y_{2}, \dots, Y_{k}) \nonumber \\
=& \frac{k+1}{2} + \frac{k+1}{2}\log(2\pi) + \frac{1}{2}\log(|\Sigma_{k+1}|) - \left( \frac{k}{2} + \frac{k}{2}\log(2\pi) + \frac{1}{2}\log(|\Sigma_{k}|) \right) \nonumber \\
=& \frac{1}{2}\left(1 + \log(2\pi) + \log\left(\frac{|\Sigma_{k+1}|}{|\Sigma_{k}|}\right) \right)
\end{align}

where $\Sigma_{k+1}$ denotes the covariance matrix for $Y_{1}, \dots, Y_{k+1}$. Note that the CI is bounded from above by the univariate entropy and obtains this value when $Y_{k+1}$ is independent of $Y_{1}, \dots, Y_{k}$.

Since clustered observations provide repeated information, we propose to downweigh these in favour of observations which provide more information on the process at other locations. To this end, the observations' CI values can be used to derive weights for weighted model calibration. 

Note that since we are using a continuous Gaussian distribution to model the process $Y$, the conditional information values may be negative and hence difficult to normalise such that they sum up to the total number of observations as suggested by \cite{Jeong2017} in their weighted calibration scheme. In addition while our motivation for weighted calibration is similar, we use CI to quantify the amount of process uncertainty reduced by the additional observation, instead of computing the uncertainty of the observation given all the other data points as done by \cite{Jeong2017}. This is because while the Jacobian they use to compute weights may be tractable in their context of systems biology, it can be difficult to calculate for complex process-based models. 

\subsubsection{Transforming the conditional information}

Rather than using trying to normalise the CI for weights, we can transform the CI values. We propose to transform the CI so that the resulting weights satisfy these conditions: they are equal to $1$ for independent points and $\frac{1}{k+1}$ for observations which lie in the same location as $k$ other existing points. The transformation is performed in three steps:

\begin{enumerate}
    \item[Step 1] \begin{equation}
        f_{1} = 1 - \exp(-2(H(Y_{k+1}) - H(Y_{k+1}|Y_{1}, Y_{2}, \dots, Y_{k})))
        \end{equation}
    \item[Step 2] \begin{equation}
         f^{*}_{1} = \frac{(m+1)(km+1)}{(k+1)m^{2}} f_{1}
         \end{equation}
    \item[Step 3] \begin{align}
    w =& 1 - f^{*}_{1}   \label{eqn:entropywts}
    \end{align}
\end{enumerate}

Here, $m$ denotes the ratio of the partial sill of the variogram to the nugget. From Step 1, we note that when $k+1$th location is far away and $Y_{k+1}$ can be seen as independent of $Y_{1}, \dots, Y_{k}$, $f_{1} = 0$ and its weight $w = 1.$ Step 2 converts $f_{1}$ to $f^{*}_{1}$ to account for the dependence of the former term on $m$ and $k$. This also enables us to obtain the desired value of $w = \frac{1}{k+1}$ when the $k+1$ observations coincide at the same location. In this scenario, the spatial covariance matrix takes the form of:

\begin{equation}
\Sigma_{k+1} = \tau^{2} \begin{pmatrix} (m+1) & m & \dots & m \\ \vdots & (m+1) & \dots & m \\ m & \dots & \ddots & m \\ m & \dots & \dots & (m+1)\end{pmatrix},
\end{equation}

where $\tau^2$ is the nugget variance. Using 2. on p.203 of of \cite{Harville1997} with $x = m$ and $\lambda = 1$, we have:

\begin{align}
|\Sigma_{k}| =& (\tau^{2})^{k}\left(km + 1\right), \nonumber \\  |\Sigma_{k+1}| = (\tau^{2})^{k+1}\left((k+1)m + 1\right). \nonumber \\ 
\Rightarrow f_{1} =& 1 - \frac{|\Sigma_{k+1}|}{|\Sigma_{k+1, k+1}||\Sigma_{k}|} \nonumber \\
=& 1 - \frac{(\tau^{2})^{k+1}\left((k+1)m + 1\right)}{(m+1)(\tau^{2})^{k+1}\left(km + 1\right)} \nonumber \\
=& 1 - \frac{(k+1)m + 1}{\left(km + 1\right)(m+1)} \nonumber \\ 
=& \frac{\left(km + 1\right)(m+1) - (k+1)m - 1}{\left(km + 1\right)(m+1)} \nonumber \\
=& \frac{km^{2}}{\left(km + 1\right)(m+1)
} \nonumber \\
\Rightarrow f^{*}_{1} =& \frac{(m+1)(km+1)}{(k+1)m^{2}}  \times \frac{km^{2}}{\left(km + 1\right)(m+1)} \nonumber \\ 
=& \frac{k}{k+1} \nonumber \\
\Rightarrow w =& 1 - \frac{k}{k+1} \nonumber \\ 
=& \frac{1}{k+1}, 
\end{align}

as required.

For the multivariate normal distribution, $f_{1}$ has the interpretation of being the proportion of variance reduced by kriging with $X_{1}, \dots, X_{k}$:

\begin{equation}
f_{1} = 1 - \exp\left(-2\left(\frac{1}{2}\log\left(\frac{|\Sigma_{k+1, k+1}||\Sigma_{k}|}{|\Sigma_{k+1}|}\right)\right)\right) = 1 - \frac{|\Sigma_{k+1}|}{|\Sigma_{k+1, k+1}||\Sigma_{k}|}
\end{equation}

If we write

\begin{equation}
\Sigma_{k+1} = \begin{pmatrix} \Sigma_{k} & \vdots \\ \dots & \Sigma_{k+1, k+1}\end{pmatrix},
\end{equation}

using (E.1) on p.204 of \cite{Harville1997}, we have:

\begin{equation}
|\Sigma_{k+1}| \\= |\Sigma_{k}|\left[|\Sigma_{k+1, k+1}| - (\sigma_{1, k+1} \dots \sigma_{k, k+1})^{T}\Sigma_{k}^{-1}(\sigma_{1, k+1} \dots \sigma_{k, k+1})\right] \\ = |\Sigma_{k}|\hat{\sigma}^{2}_{k+1|1, \dots, k},
\end{equation}

where $\sigma_{i, k+1}$ denotes the $(i, k+1)$th entry of $\Sigma_{k+1}$ and $\hat{\sigma}^{2}_{k+1|1, \dots, k}$ denotes the (simple) kriging variance when we use $Y_{1}, \dots, Y_{k}$ to predict $Y_{k+1}$ \citep{Chiles1999}. 

So:

\begin{equation}
f_{1} = \frac{|\Sigma_{k+1, k+1}| - \hat{\sigma}^{2}_{k+1|1, \dots, k}}{|\Sigma_{k+1, k+1}|},
\end{equation}

ie. the proportion of variance reduced by kriging with $Y_{1}, \dots, Y_{k}$. Notice that when $Y_{k+1}$ is independent of $Y_{1}, \dots, Y_{k}$, $f_{1} = 0 \Rightarrow w = 1$  because the kriging variance is equal to the sill.

To use the normalised CI or the weights defined in (\ref{eqn:entropywts}) for model calibration, we can proceed as follows:

\begin{enumerate}
    \item Assign equal weight to all data points.
    \item Optimise the parameters using an initial unweighted cost function.
    \item Check for spatial dependence in the model residuals and estimate the variogram if present. 
    \item Compute the weights for each data point via (\ref{eqn:entropywts}). 
    \item Use the weights in the next round of model calibration via the weighted cost functions.
\end{enumerate}

If necessary, we can iterate the calibration and computation of weights until the weights converge as suggested by \cite{Jeong2017} in the context of weighted calibration for systems biology. However, in majority of the cases we analysed, the difference between weights computed in the first and second iterations were very small and no further iteration was required.
\clearpage

\section{Simulation experiments using Gaussian processes} \label{sec:GPsim}

To test the effect of our proposed weighting scheme, we apply it to simulated datasets. Specifically, we simulate values from Gaussian processes to generate spatial dependence. To imitate the modelling framework for the tephra case study, the means of the processes are set to zero because deterministic long-range trends will be captured by the process-based model instead of a non-zero mean and/or regression term. In addition, we simulate over a spatial domain of $[0, 50] \times [0, 50]$ to mimic 50 km by 50 km area covered by our tephra load dataset. 

 By comparing estimates of the mean obtained from minimising weighted and unweighted MSE for different simulation settings, we can examine the properties of weighted calibration. $100$ independent datasets are generated to evaluate each simulation setting and make comparisons. In their comparison of ordinary least squares (OLS) and generalised least squares (GLS) regression where spatial correlation was explicitly modelled, \cite{Beale2007} simulated spatially correlated data over a grid and showed that while OLS and GLS both gave an unbiased results for the slope of a linear model, GLS provided higher precision: the estimates exhibited a smaller spread about the true value. We are expecting to see similar results for our weighted calibration. In addition to examining how the benefit of weighting changes with the level of spatial dependence present in the data, we explore different sampling schemes to generate point observations which are more in line with what one would obtain in scientific field surveys.

\subsection{Effect of sampling schemes}

\begin{figure}[tbp]
    \centering
        \includegraphics[width = 0.95\textwidth]{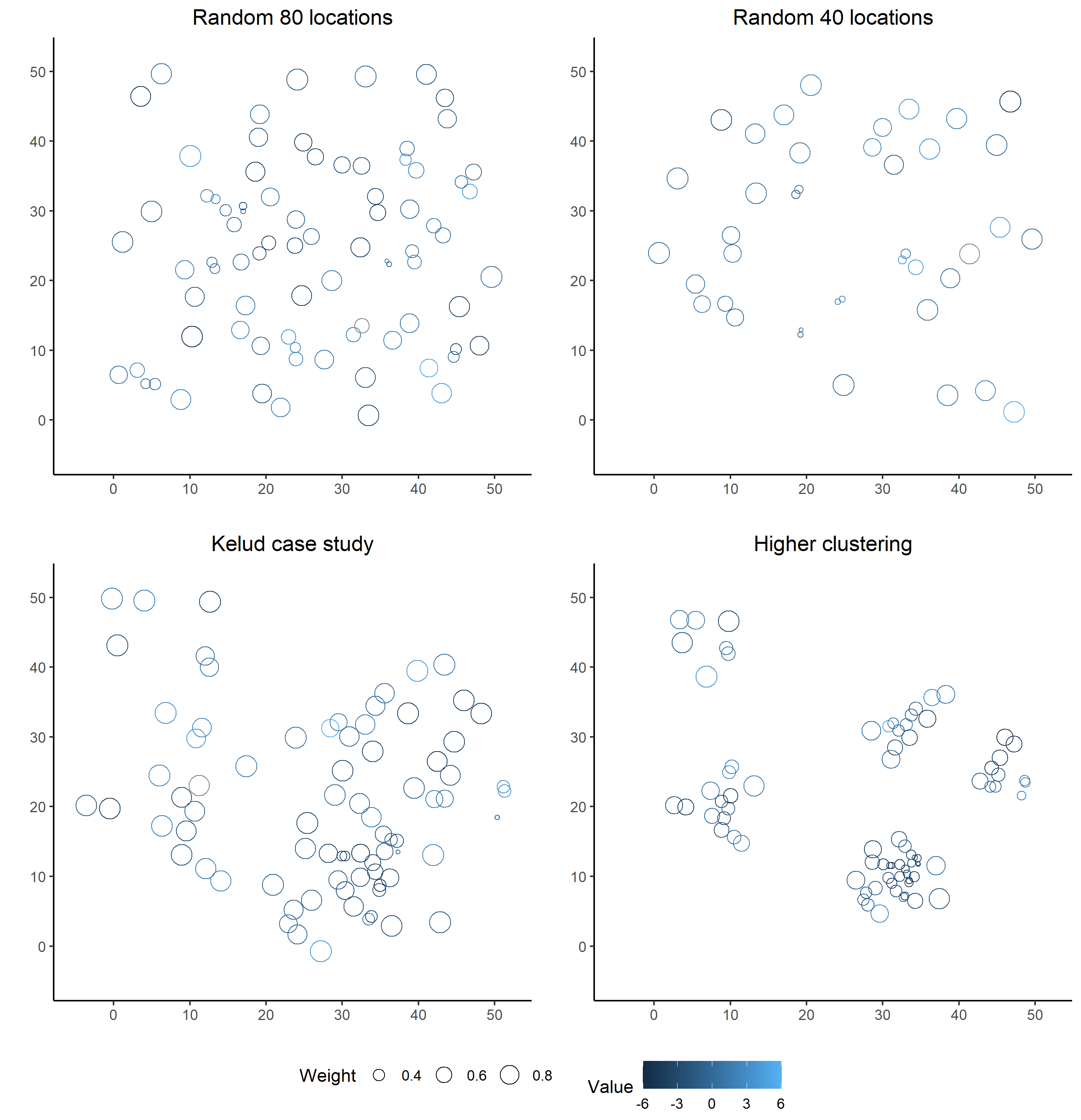}
\caption{Four different sampling schemes together with their simulated process values and resulting spatial weights. Note that the weights were computed based on the estimated variogram from the simulations separately and hence are not comparable between simulations.}
\label{fig:sampling_schemes}
\end{figure}

\begin{figure}[tbp]
    \centering
        \includegraphics[width = 0.85\textwidth]{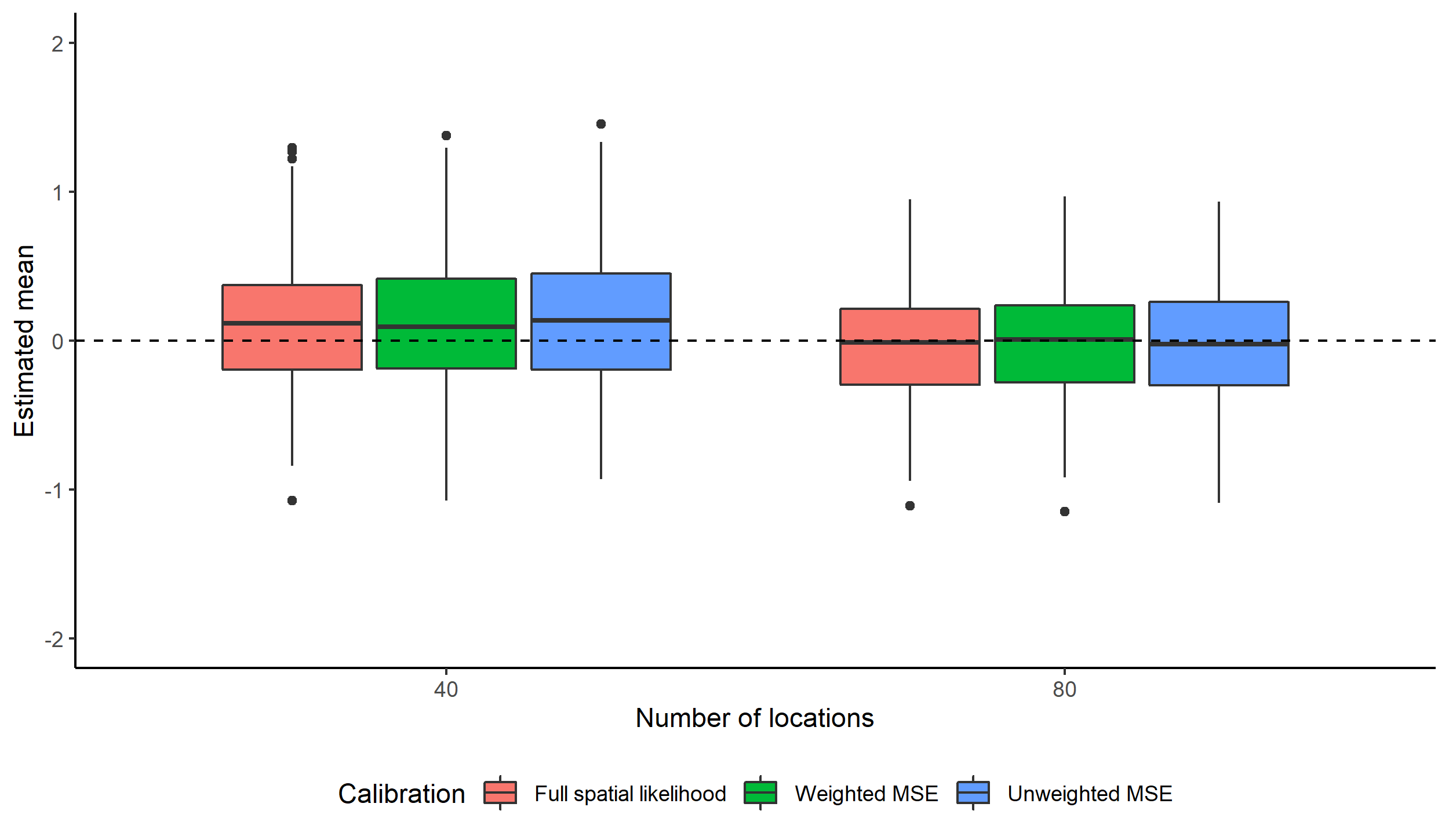}
\caption{Box plots of the estimates for the Gaussian process mean from the 100 simulations at 40 and 80 randomly sampled locations. The dashed horizontal line denotes the true mean of zero.}
\label{fig:ss_bp}
\end{figure}

\begin{figure}[bp]
    \centering
        \includegraphics[width = 0.85\textwidth]{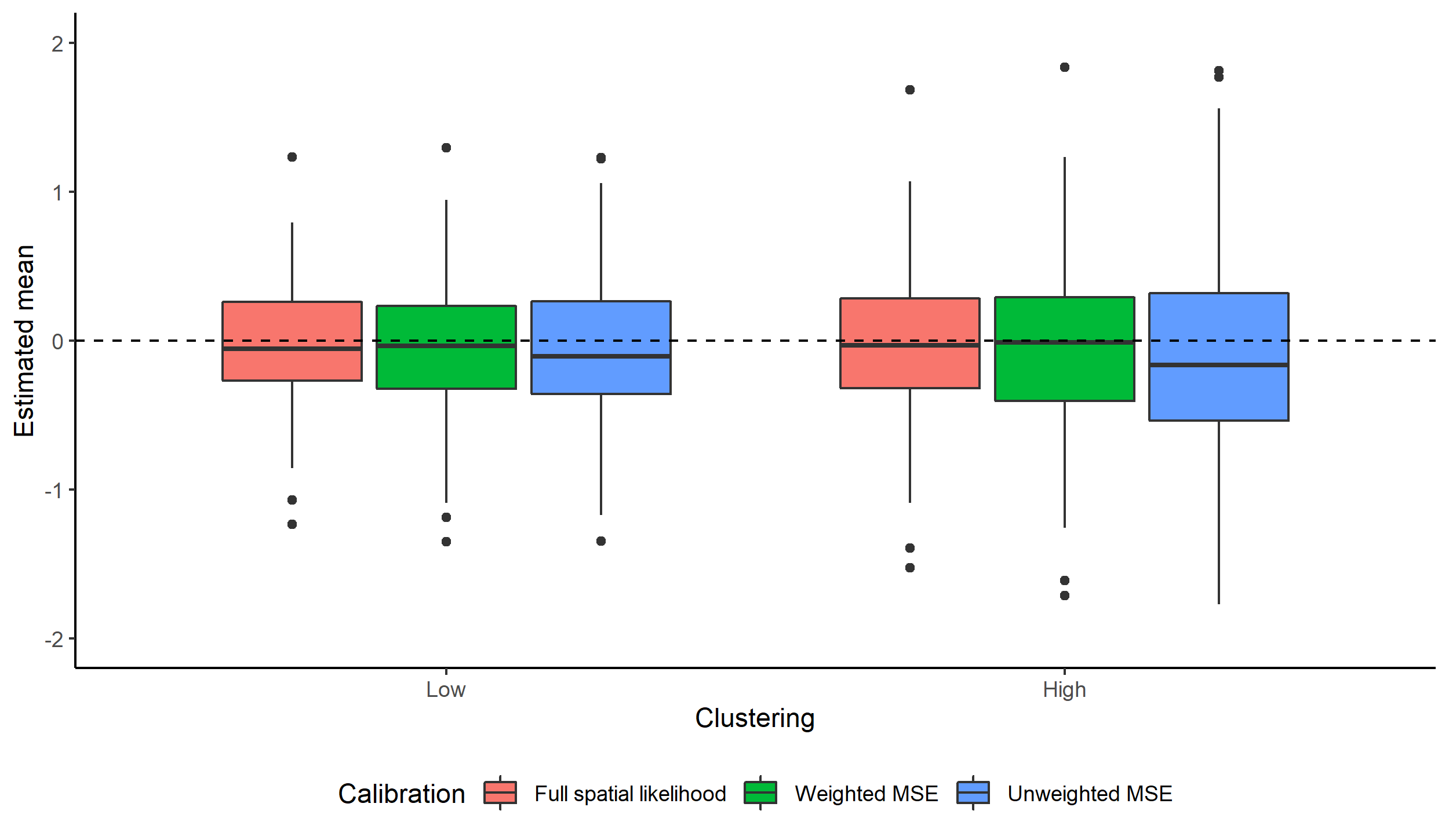}
\caption{Box plots of the estimates for the Gaussian process mean from the 100 simulations at the original Kelud case study observation locations and more clustered locations. The dashed horizontal line denotes the true mean of zero.}
\label{fig:clustering_bp}
\end{figure}

We consider both random spatial sampling and clustered sampling. In line with our tephra data, our base setting involves 80 observations for each simulation. Figure \ref{fig:sampling_schemes} shows the four sampling schemes we consider: 80 randomly chosen locations, 40 randomly chosen locations, 80 locations with relative positions fixed to be the same as that in the tephra load data from the 2014 Kelud eruption and 80 locations exhibiting higher clustering than our case study data. To generate the latter, we used k-means clustering to divide the original locations into five clusters (five being the number chosen by the elbow method) and halved the distance of each observation to its associated cluster centre. 
\\
To generate Gaussian process values with spatial dependence at the chosen locations, we use a Mat\'ern variogram with nugget $\tau^2 = 1$, sill $\sigma^{2} = 4$ and range parameter $\phi = 2.5$. These relative values are similar to what we obtain for the case study in Section \ref{sec:casestudy}. Note that the smoothness parameter $\kappa$ is set to $1$ which is a common choice which allows for slightly smoother surfaces than an exponential and the default value in the R package R-INLA \citep{rinla, rue2009}. We generate $100$ sets of simulated values per sampling scheme and estimate the process mean for each by optimising the full spatial likelihood (via the `likfit' function of the geoR package) as well as weighted and unweighted MSE.
\\
Figure \ref{fig:ss_bp} shows the box plots of the process mean estimates from the 40 and 80 randomly sampled locations for the three calibration methods. With less observations, there is higher bias and spread in the estimates as seen in the larger differences between the bold line denoting the median values from the dashed line denoting the true mean of zero as well as the longer widths of the box plots. While the WMSE calibration has slightly lower bias in the case with 40 observations than the other two methods, its benefit is hard to ascertain with 80 observations. This is likely because, by random sampling, we have a good representation of the study region and process values, and this is even more so with more observations considered.
\\
Holding the number of observations fixed at 80, we examine the effect of clustering our observations. Figure \ref{fig:clustering_bp} shows the box plots of the process mean estimates from the 80 locations corresponding to those of the Kelud case study and after increasing the extent of clustering by halving the distance of the observations to their cluster centres. Under increased clustering, we see that the bias of the unweighted calibration increases while the estimates from the weighted calibration remain unbiased. The precision of all three methods decrease with increased clustering but based on the width of the boxes, unweighted calibration is affected the most. The same conclusions can be drawn when we compare the box plots corresponding to the highly clustered observations to that for 80 randomly selected locations in Figure \ref{fig:ss_bp}.

\subsection{Extent of spatial dependence}

\begin{figure}[tbp]
    \centering
        \includegraphics[width = \textwidth]{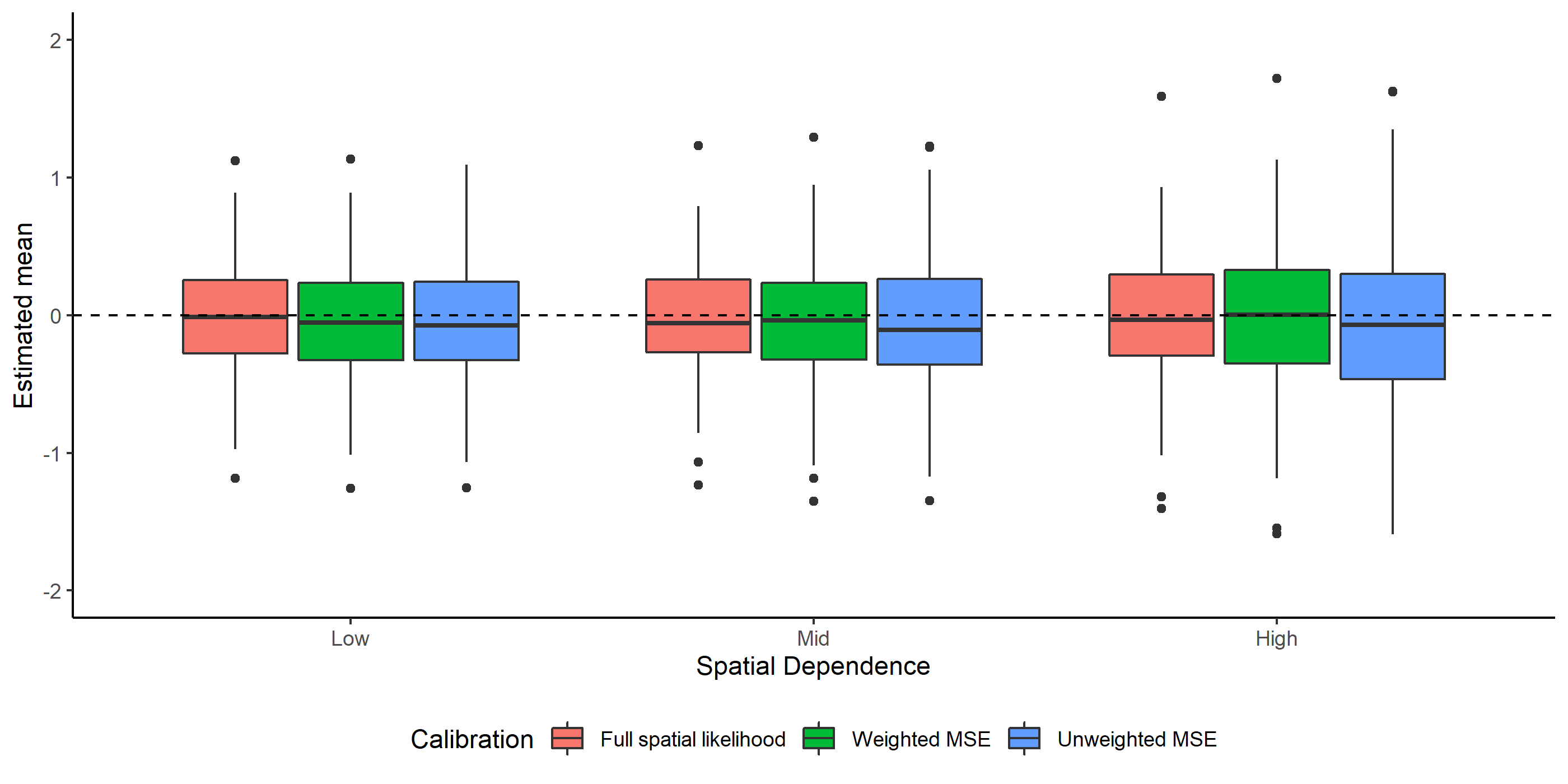}
\caption{Box plots of the estimates for the Gaussian process mean from the 100 simulations at the original Kelud case study observation locations for different spatial dependence settings. `Mid' refers to the case where the original variogram parameters are used, `Low' refers to the case where the proportion of the nugget to the sill is $0.4$ instead of $0.2$ and `High' refers to the case where the range parameter is $3.5$ instead of $2.5$. The dashed horizontal line denotes the true mean of zero.}
\label{fig:spatdep_bp}
\end{figure}

From the previous section, we saw that weighted calibration is important for clustered observations. In this section, we vary the spatial dependence of the Gaussian process to see its effect on the mean estimates. We do this by increasing the proportion of the nugget variance to the sill from $0.2$ to $0.4$ (hence increasing random noise and reducing spatial dependence from the `Mid'/base case) and increasing the spatial range parameter $\phi$ from $2.5$ to $3.5$ (hence increasing dependence). In Figure \ref{fig:spatdep_bp}, the former case is represented by `Low' and the latter by `High'. Generally, we see that when we increase the spatial dependence in the Gaussian process, the precision of the mean estimates decreases (the boxes are wider). Nevertheless, the estimates from the weighted calibration remain relatively unbiased under Mid-High spatial dependence unlike those from the unweighted calibration and exhibit higher precision.

\clearpage

\section{Weighted calibration of Tephra2} \label{sec:Tephra2}

In Section \ref{sec:GPsim}, we saw that for the simulated Gaussian processes, the benefits of weighted calibration over unweighted calibration in terms of lower bias and higher precision increase with observation clustering and spatial dependence. Now, we apply our methods to the calibration of the Tephra2 dispersion model with tephra load data collected in the aftermath of the 2014 Kelud eruption. We show that there are non-negligible differences between the estimated source parameters from weighted and unweighted calibration, and use simulation experiments to further illustrate the lower bias and higher precision obtained from weighted calibration with this process-based model. 

\subsection{Application: 2014 eruption of the Kelud volcano} \label{sec:casestudy}

\begin{figure}[tbp]
    \centering
        \includegraphics[width = 0.75\textwidth]{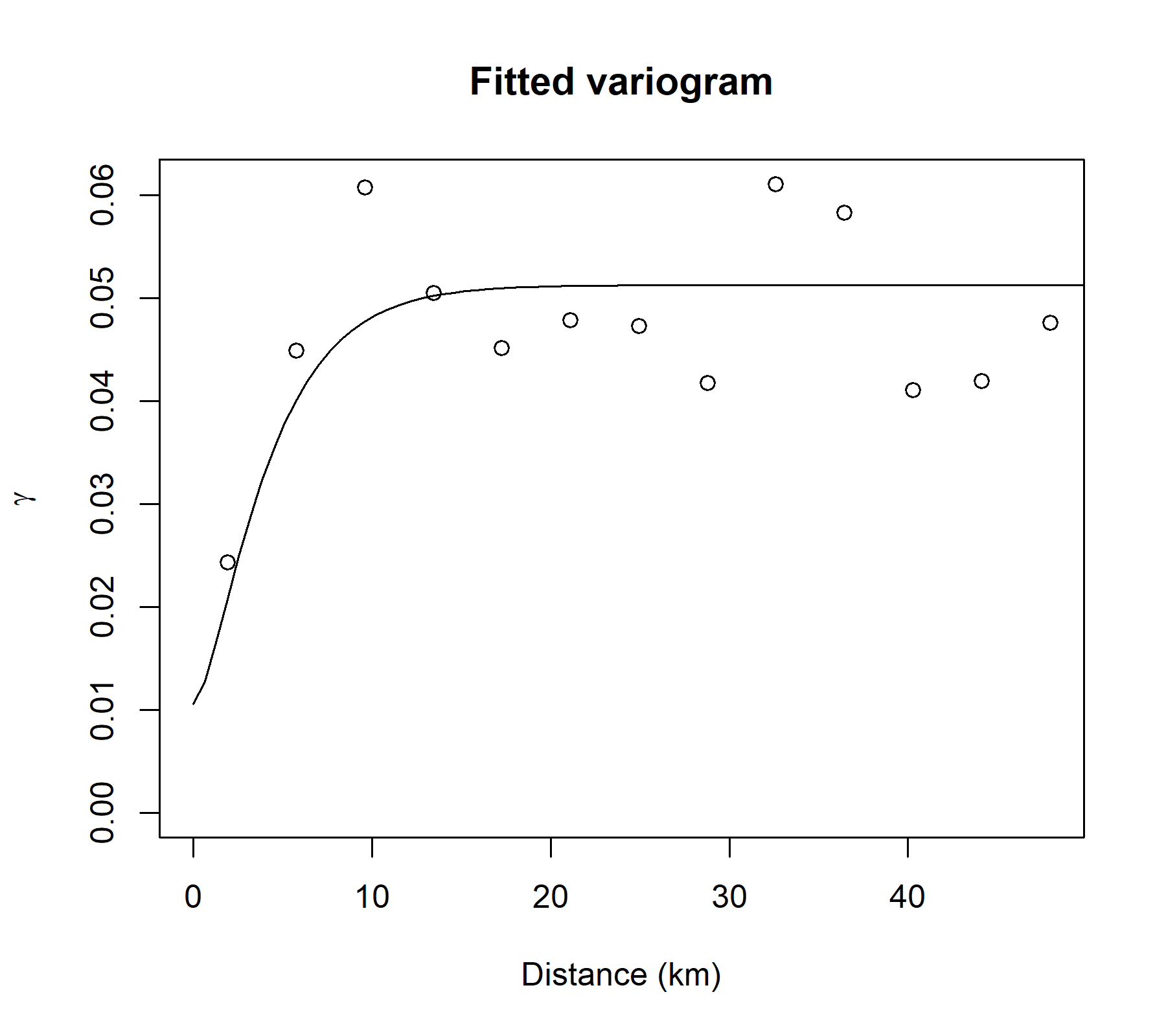}
\caption{Mat\'ern fit (black line) to the empirical variogram (points) of the Tephra2 model residuals for the 2014 Kelud case data. Here, the nugget variance $\tau^2 = 0.0105742$, the partial sill $\sigma^{2} = 0.04067892$ and the range parameter $\phi = 2.84104255$ with a practical range of $11.360$km where correlation reaches 5\%. The smoothness parameter $\kappa$ is fixed at 1.}
\label{fig:kelud_vgm}
\end{figure}

\begin{figure}[tbp]
    \centering
        \includegraphics[width = 0.9\textwidth]{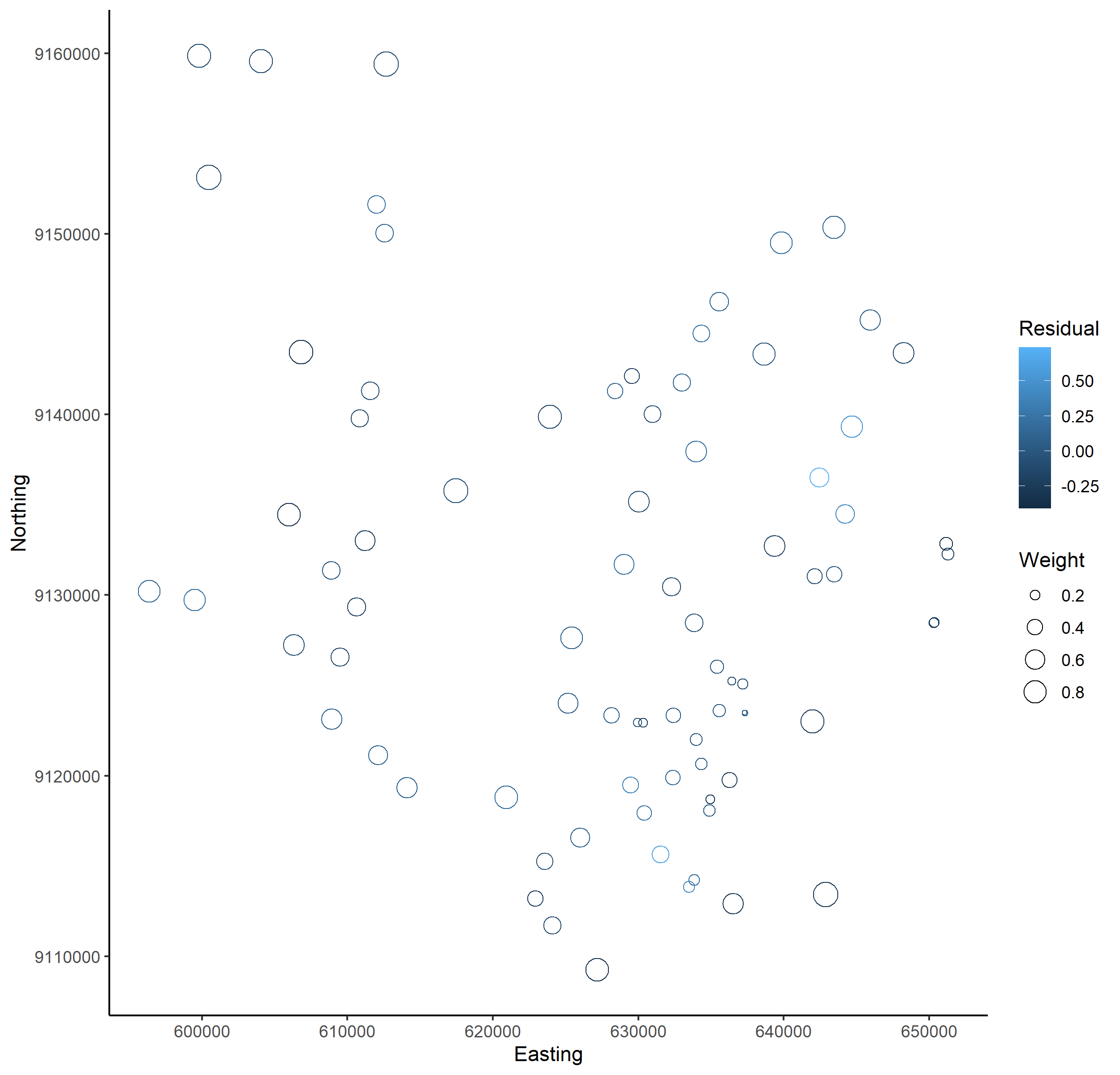}
\caption{Spatial map of the observation locations with their associated Tephra2 model residuals and computed weights.}
\label{fig:weights_parall}
\end{figure}

\begin{figure*}[t!]
    \centering
    \begin{subfigure}[t]{0.5\textwidth}
        \centering
        \includegraphics[width = \textwidth]{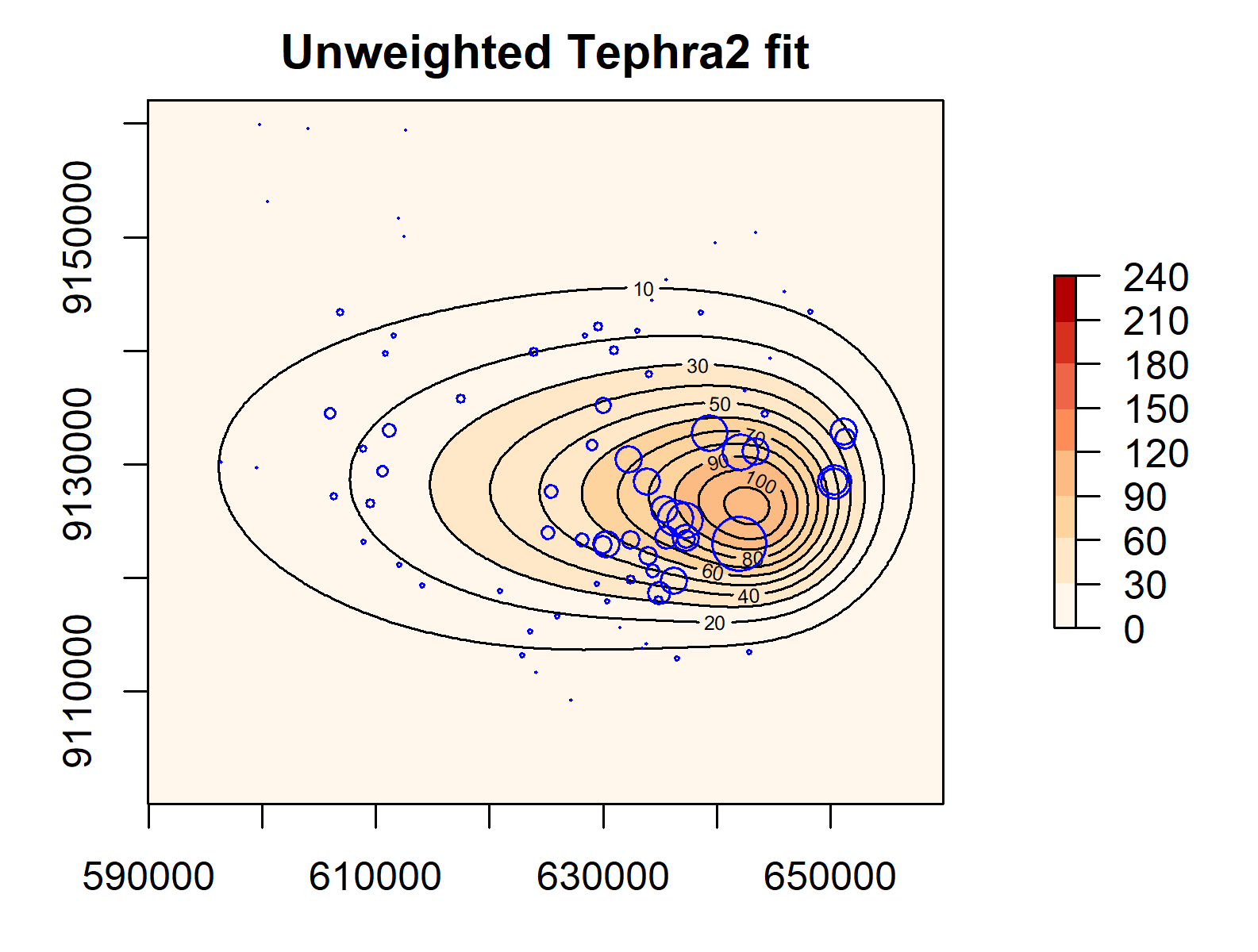}
        \caption{}
    \end{subfigure}%
    ~ 
    \begin{subfigure}[t]{0.5\textwidth}
        \centering
        \includegraphics[width = \textwidth]{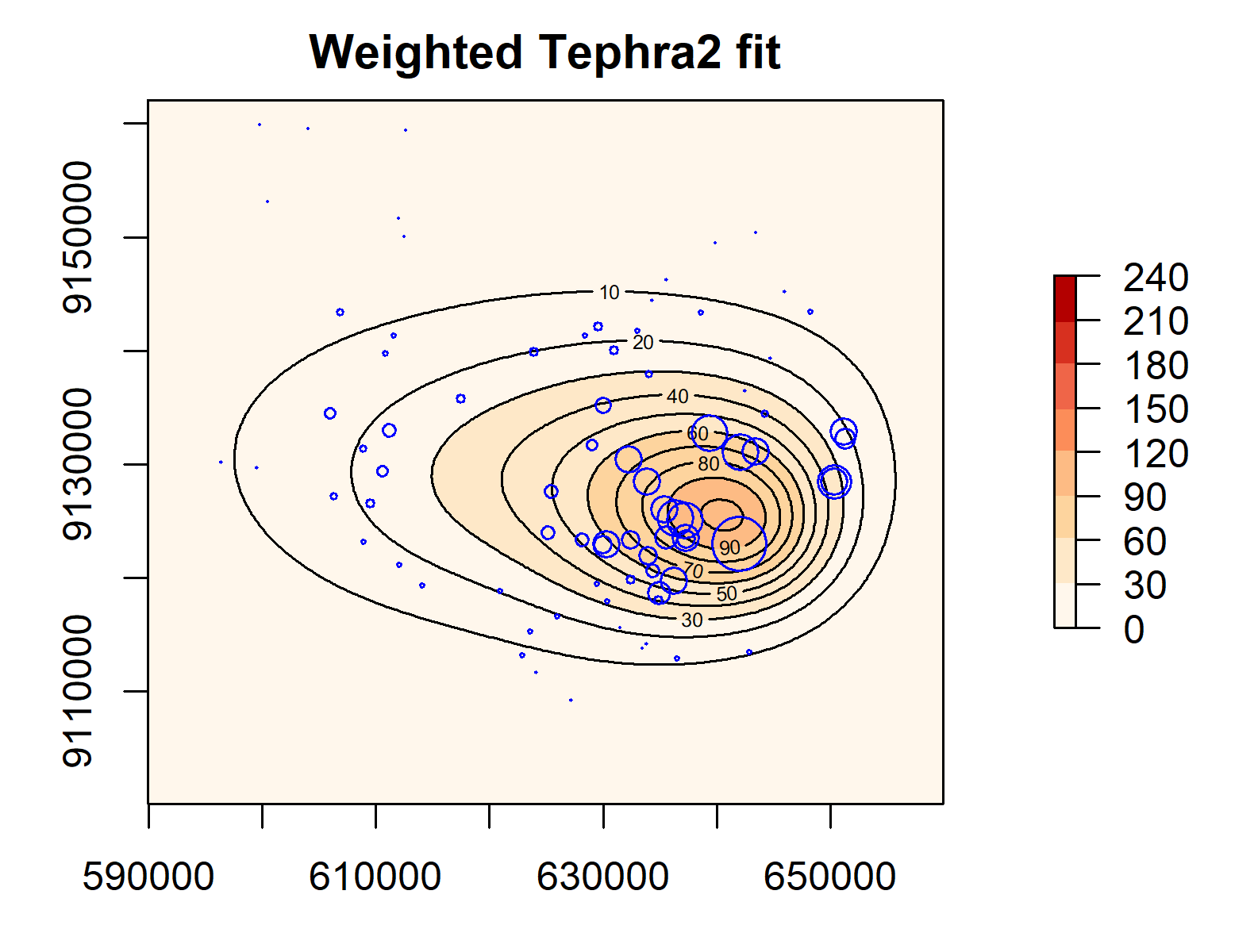}
        \caption{}
    \end{subfigure}
        \caption{Spatial maps of estimated tephra load in kg/m$^2$ from (a) the unweighted calibration and (b) the weighted calibration.}
        \label{fig:grid_kelud}
\end{figure*}

We conduct weighted and unweighted calibration of the Tephra2 model using the tephra load data from the 2014 Kelud eruption. Note that calibration was not conducted using a full spatial likelihood unlike in the simulation experiements for the Gaussian process in Section \ref{sec:GPsim} because of the infeasibility of this approach for large environmental datasets. The fitted variogram for the model residuals from the unweighted calibration is shown in Figure \ref{fig:kelud_vgm}. The increase from a nugget value of $\tau^2 = 0.0106$ to the sill of $0.0512$ indicates that there is spatial correlation present since the expected difference in residuals increases with increasing distance between observations. 
\\
We use the fitted variogram to construct the spatial covariance matrix required to compute the CI weights and conduct the weighted calibration. The computed weights are shown on a map in Figure \ref{fig:weights_parall}. We see that clustered observations are given lower weights. Table \ref{tab:kelud_parall} shows the parameter estimates from the weighted and unweighted calibration. We see that the difference in estimates ranges from -71.587\% for fall time threshold to +59.295\% for the standard deviation in particle diameter. Hence, the effect of accounting for spatial dependence through weighting is non-negligible. As expected, the difference in total mass ejected is relatively low at +3.364\% as it is known to be the parameter which is most independent of the rest and easiest to estimate \textit{(ref from Seb?)}.  
\\
The difference in fitted parameters manifests in the estimated tephra load surface as illustrated in Figure \ref{fig:grid_kelud}. The contours from the weighted calibration, denoting areas of similar load values, are less elongated towards the left than those from the unweighted calibration. In addition, there is generally lower loads at distal and medial regions (between the 30-100 kg/m$^2$ contours).

\clearpage

\begin{table}[tbp]
\begin{tabular}{l|lllr}
\multirow{2}{*}{\textbf{Parameter}}            & \multirow{2}{*}{\textbf{Range}} & \textbf{Unweighted}  & \textbf{Weighted}    & \multirow{2}{*}{\textbf{Difference}}        \\
                                               &                                 & \textbf{calibration} & \textbf{calibration} &  \\
                                               \hline \\
Max column height (km a.s.l.)         & 15 - 23                         & 22.718               & 17.966               & - 20.917\%          \\
Total mass ejected ($10^{10}$ kg)                & 0.1 - 10                        & 6.61292              & 6.83535              & + 3.364\%            \\
Diffusion coefficient (m\textasciicircum{}2/s) & 0.0001 - 15000                  & 10403                & 13823                & + 32.875\%           \\
Beta                                           & 0.001 - 3.5                     & 2.7677               & 1.25203              & - 54.763\%           \\
Fall time threshold (s)                        & 0.001 - 15000                   & 11382                & 3234                 & - 71.587\%           \\
Median grain size (phi)                        & -3 - 2.5                        & 0.859525             & 0.877705             & + 2.115\%            \\
SD in particle diameter (phi)           & 0.5 - 3                         & 0.840127             & 1.33828              & + 59.295\%          
\end{tabular}
\caption{Difference in Tephra2 model parameter estimates from the weighted and unweighted calibration, together with the parameter ranges explored in the optimisation. Here, a.s.l. refers to above sea level and SD refers to standard deviation.}
\label{tab:kelud_parall}
\end{table}

\clearpage 

\subsection{Simulation experiments}

\begin{figure*}[tbp]
    \centering
    \includegraphics[width = \textwidth]{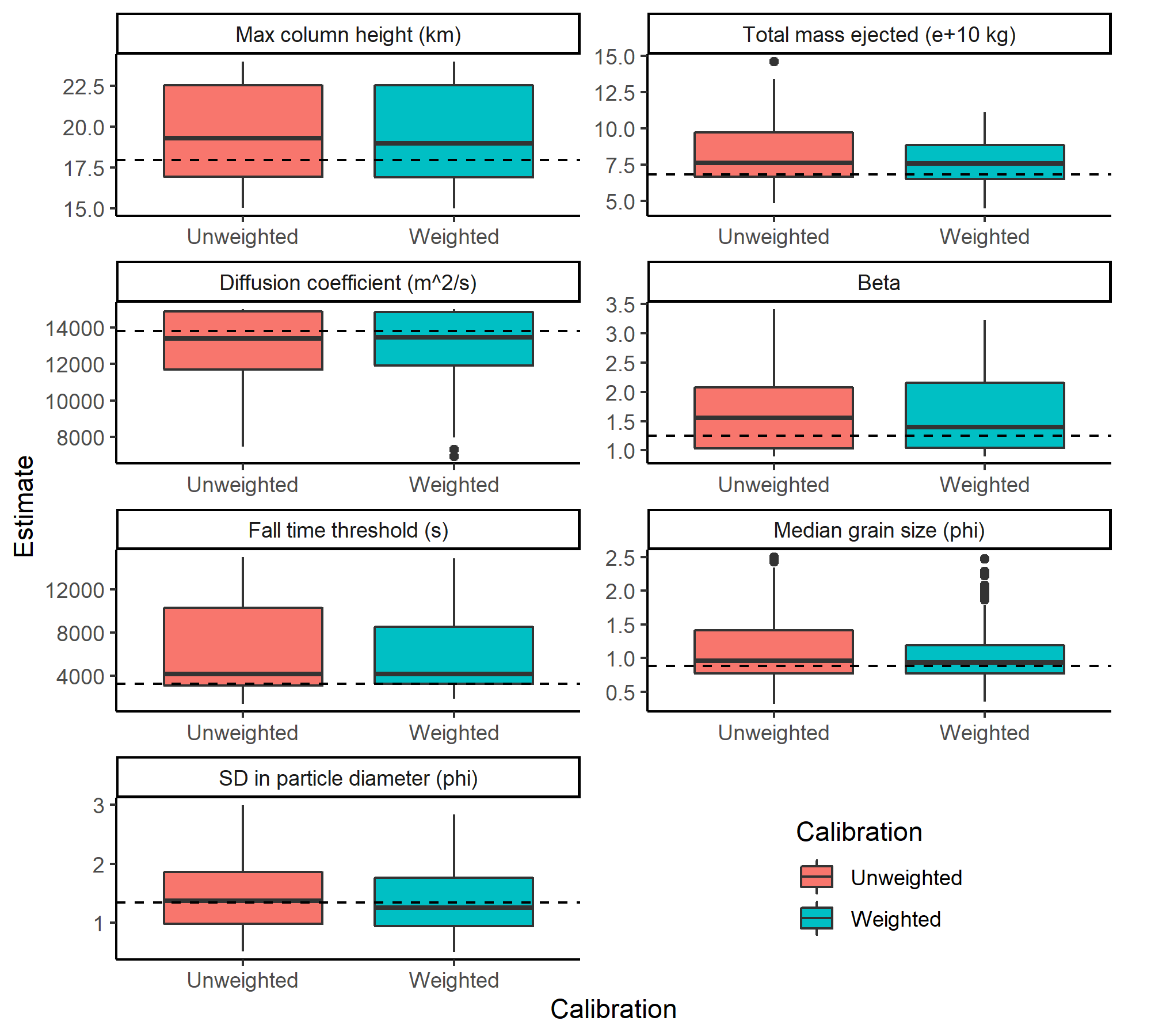}
     \caption{Parameter estimates of the Tephra2 model from weighted and unweighted calibration. The horizontal dashed lines denote the true values of the parameters.}
     \label{fig:sim_par_est}
\end{figure*}

\begin{figure*}[tbp]
    \centering
    \includegraphics[width = \textwidth]{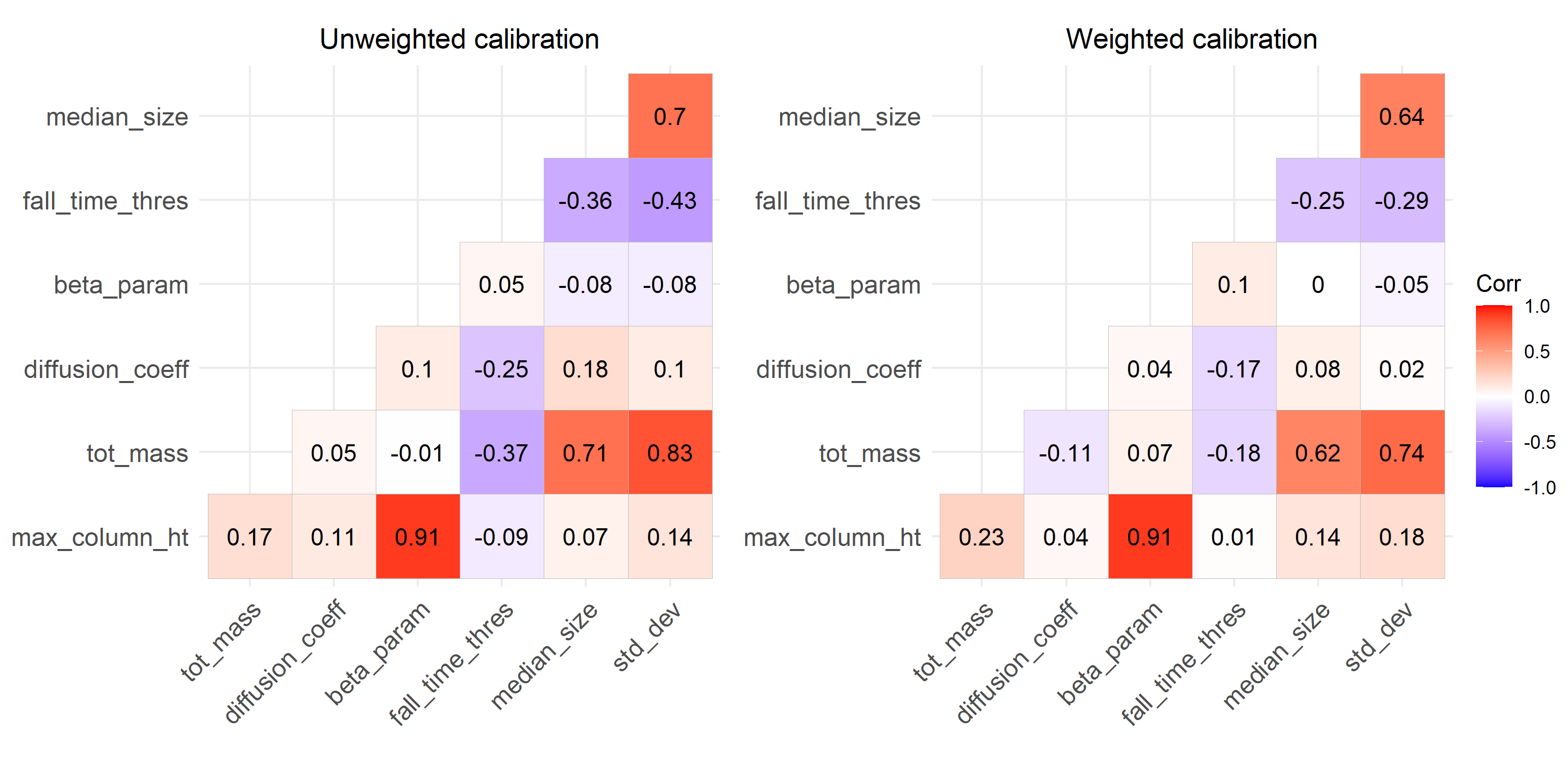}
     \caption{Estimated correlation between the parameter estimates of the Tephra2 model from unweighted calibration.}
     \label{fig:par_corr}
\end{figure*}

\begin{figure*}[tbp]
    \centering
    \includegraphics[width = 0.7\textwidth]{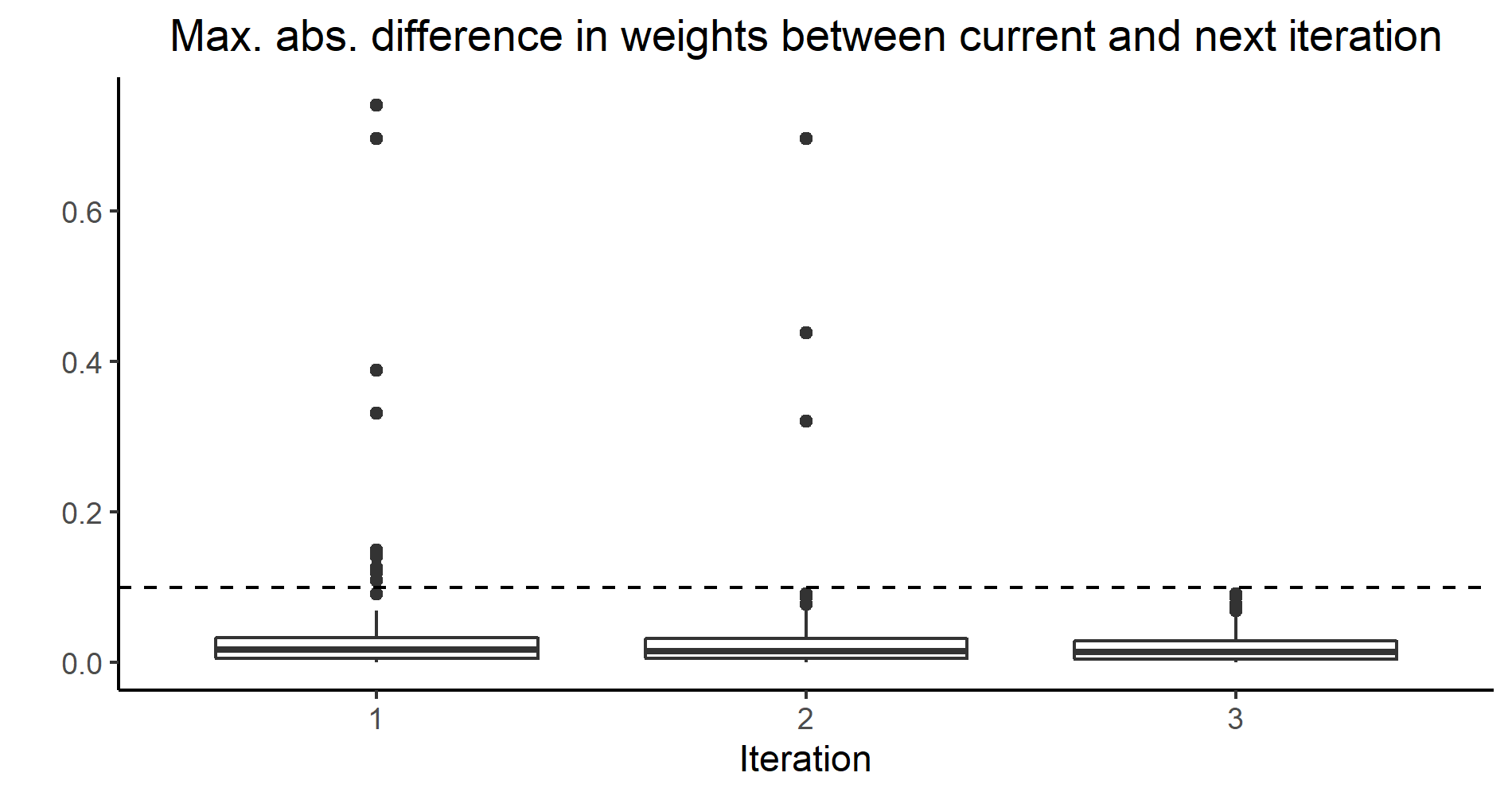}
     \caption{Maximum absolute difference in weights between consecutive calibration iterations. The dashed horizontal line denotes a chosen threshold of $0.1$ to determine if reweighting is needed.}
     \label{fig:wt_diff}
\end{figure*}

With the estimated variogram parameters (nugget, partial sill and range), we can generate 100 sample tephra load datasets from the same distribution as the 2014 Kelud data by adding simulated residual data to the Tephra2 fitted values from weighted inference. This means that we treat the Tephra2 fit as a mean surface of tephra load and use the Gaussian process with the fitted variogram to generate spatially correlated residuals. Examples of two simulated tephra load datasets are shown in Figure \ref{fig:tephra_sim} of the Appendix.
\\
We perform weighted and unweighted calibration of the Tephra2 model for each of these 100 simulated datasets to examine the differences in the inferred parameters. Figure \ref{fig:sim_par_est} shows the box plots of the parameter estimates from the two calibration methods and Table \ref{tab:keludsim_parall} summarises the results in terms of percentage median bias (the difference between the median estimate and the true parameter value expressed in terms of percentage of the latter) as well as percentage median absolute deviation, a measure of spread. From these, we see that weighted calibration clearly improves inference of certain parameters. The median bias for beta parameter and median grain size decreases from 23.86\% to 11.94\% and from 8.97\% to 6.56\% respectively. In addition, their median absolute deviation decreases from 61.93\% to 55.15\% and from 41.95\% to 35.31\% respectively. Apart from the diffusion coefficient (for which accuracy and precision seems unaffected) and the standard deviation in particle diameter (for which estimates slight increase in absolute bias), weighting based on spatial conditional information either improves accuracy or precision for the remaining Tephra2 parameters. The median absolute deviation of total mass ejected decreases from 30.61\% to 25.37\% while that for fall time threshold decreases from 83.23\% to 65.60\%. On the other hand, maximum column height sees a reduction in median bias from 7.46\% to 5.75\%. Note that the procedure of estimating parameters from data simulated from the fitted model can be used to construct parametric bootstrap confidence intervals for the parameters. In the cases where the median absolute deviation is lower for estimates from the weighted calibration, the corresponding confidence intervals will be narrower than those derived from unweighted calibration.
\\
The unequal effects of weighting on the different parameters is likely due to their interactions in the physical model. Figure \ref{fig:par_corr} shows that there are some high correlations between parameters estimates from both unweighted and weighted calibration. In particular, standard deviation in particle diameter (std\_dev) is highly correlated to median grain size (median\_size) and total mass ejected (tot\_mass) and the beta parameter (beta\_param) is highly correlated with the maximum column height (max\_column\_ht).
\\
Figure \ref{fig:wt_diff} shows the maximum absolute difference in weights between consecutive calibration iterations. Using a convergence threshold of 0.1 to determine if additional iterations are required, 90 out of the 100 datasets had weights which converged after one round of weighting. Only 6 out of the 100 datasets required two rounds of weighting for the weights to converge while 4 out of the 100 datasets required three rounds. 

\clearpage

\begin{table}[tbp]
\begin{tabular}{l|cc|cc}
\multirow{2}{*}{\textbf{Parameter}} & \multicolumn{2}{c|}{\textbf{ \% Median-bias}}   & \multicolumn{2}{c}{\textbf{\% Median abs. deviation}} \\
& \textbf{Unweighted} & \textbf{Weighted} & \textbf{Unweighted}  & \textbf{Weighted} \\ \hline & \multicolumn{2}{c|}{} & \multicolumn{2}{c}{} \\
Max column height & \textbf{7.46\%} & \textbf{5.75\%} & 22.95\% & 22.47\% \\
Total mass ejected & 11.38\% & 10.80\% & \textbf{30.61\%} & \textbf{25.47\%} \\
Diffusion coefficient &   -2.93\% &  -2.68\%  & 16.01\% & 15.43\%                          \\
Beta  &  \textbf{23.86\%} & \textbf{11.94\%}  & \textbf{61.93\%} & \textbf{55.14\%} \\
Fall time threshold  & 29.11\%  & 28.39\% & \textbf{83.23\%} & \textbf{65.60\%}                        \\
Median grain size  &  \textbf{8.97\%} & \textbf{6.56\%} & \textbf{41.95\%} & \textbf{35.31\%}                       \\
SD in particle diameter &  \textbf{2.57\%} &  \textbf{-6.52\%} &   \textbf{48.84\%} & \textbf{46.52\%}          
\end{tabular}
\caption{Percentage difference between the median estimates and the true parameter values (\% median-bias), as well as the percentage median absolute deviation of the estimates from the weighted and unweighted calibrations of the Tephra2 model for $100$ spatially-correlated simulations. The numbers are in bold if greater than 1\% in absolute terms.}
\label{tab:keludsim_parall}
\end{table}

\clearpage

\clearpage

\bibliographystyle{agsm} 
\renewcommand\refname{REFERENCES}

\bibliography{refs}

\section*{Appendix}

\subsection*{Weighted calibration in systems biology}

In the context of system biology, a similar problem is faced where limited experimental data is used to fit complex models and not all data points provide equal amounts of information on the system. To address this, \cite{Jeong2017} propose a calibration approach, where more informative data points were given higher weights. In their case, these were data points taken during the dynamic phase of the experiment rather than those taken during the steady state. 
\\
In their paper, \cite{Jeong2017} consider the weighted sum of squares as a cost function:

\begin{equation}
    C(\theta) = \frac{1}{2}\sum\limits_{i = 1}^{n}\left(Y_{i}^{obs} - Y_{i}^{pred}\left(\theta\right)\right)^{2} w_{i}
\end{equation}

where $\theta$ represents the model parameters, $Y_{i}^{obs}$ the $i^{th}$ observation, $Y_{i}^{pred}$ the corresponding prediction and $w_{i}$ its allocated weight. Through a second-order Taylor expansion of $C$ centered around the optimal parameters and some simplifications, parameter uncertainty can be expressed in terms of the Fisher information matrix. In a similar way, the uncertainty of estimating one set of data points using another set of data points can be expressed in terms of their Fisher information matrices.
\\
The iterative algorithm is as follows:

\begin{enumerate}
    \item Assign equal weight to all data points and use a random initial parameter setting.
    \item Optimise the parameters using an initial unweighted cost function.
    \item Compute Jacobian $J$ at the optimised parameter values.
    \item Compute the uncertainty of data point $i$ given all the other data points as:
    \begin{equation}
    U(i | S_{-i}) = \frac{1}{m}trace(I_{i}I_{S_{-i}}^{-1})
    \end{equation}
    where $m$ is the number of parameters, $I_{i} = J_{i}^{T}J_{i}$ where $J_{i}$ is the $i^{th}$ row of $J$ and $I_{S_{-i}}$ is computed using all the other rows of $J$. The term for which we compute trace approximates the derivatives of the other data points with respect to $i$ \cite[p.42]{Jeong2019}.
    \item Compute the weights $w_{i}$ by normalising the uncertainty values so that they sum up to the total number of data points $n$.
    \item Use the weights in the next round of model calibration via the weighted cost function and use the previously estimated parameters as initial parameter settings.
    \item Iterate the calibration and and computing of weights until the weights converge.
\end{enumerate}

\subsection*{Examples of simulated tephra data}

\begin{figure*}[t!]
    \centering
    \begin{subfigure}[t]{0.75\textwidth}
        \centering
        \includegraphics[width = \textwidth]{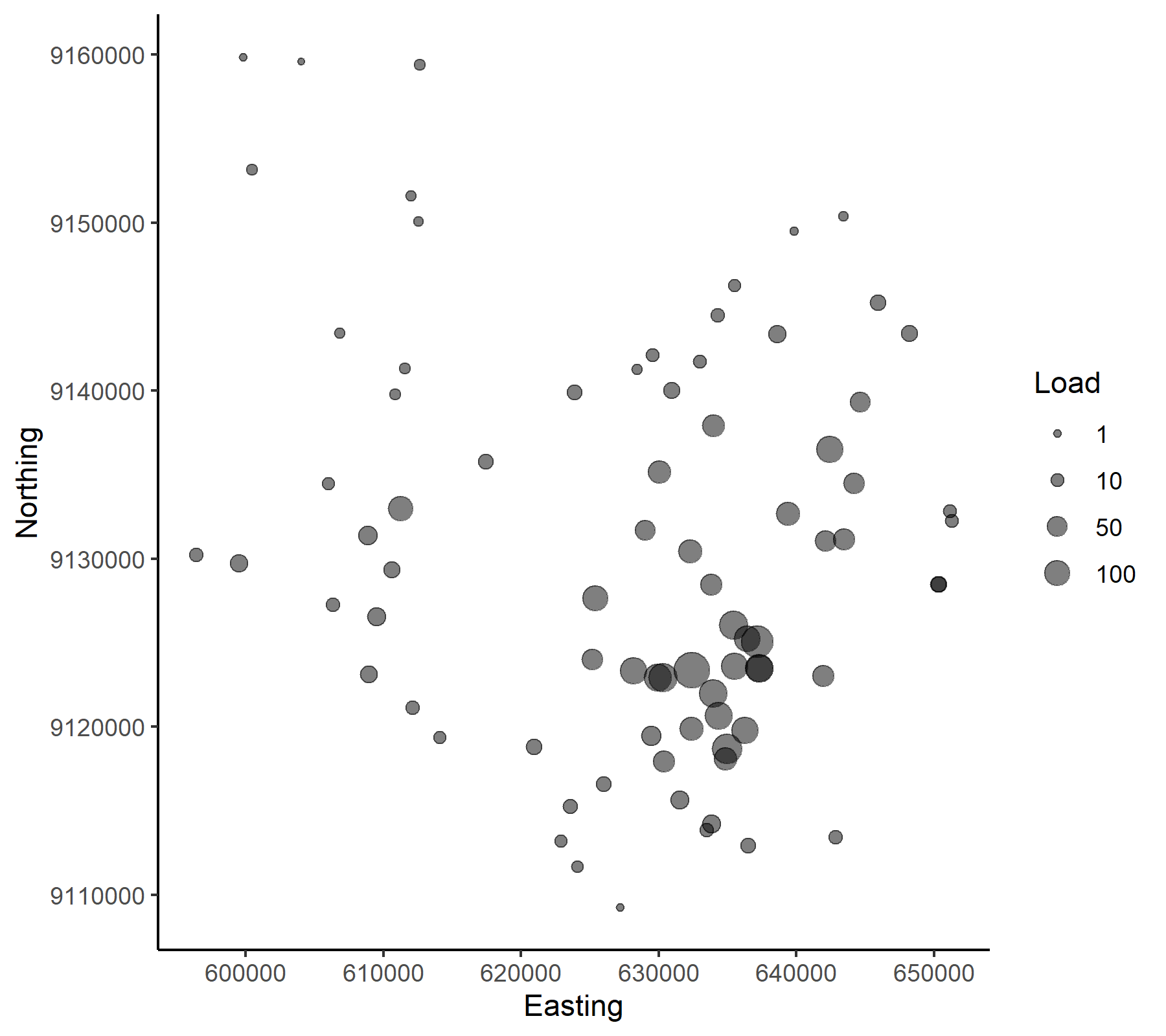}
    \end{subfigure}%
   \hfill
    \begin{subfigure}[t]{0.75\textwidth}
        \centering
        \includegraphics[width = \textwidth]{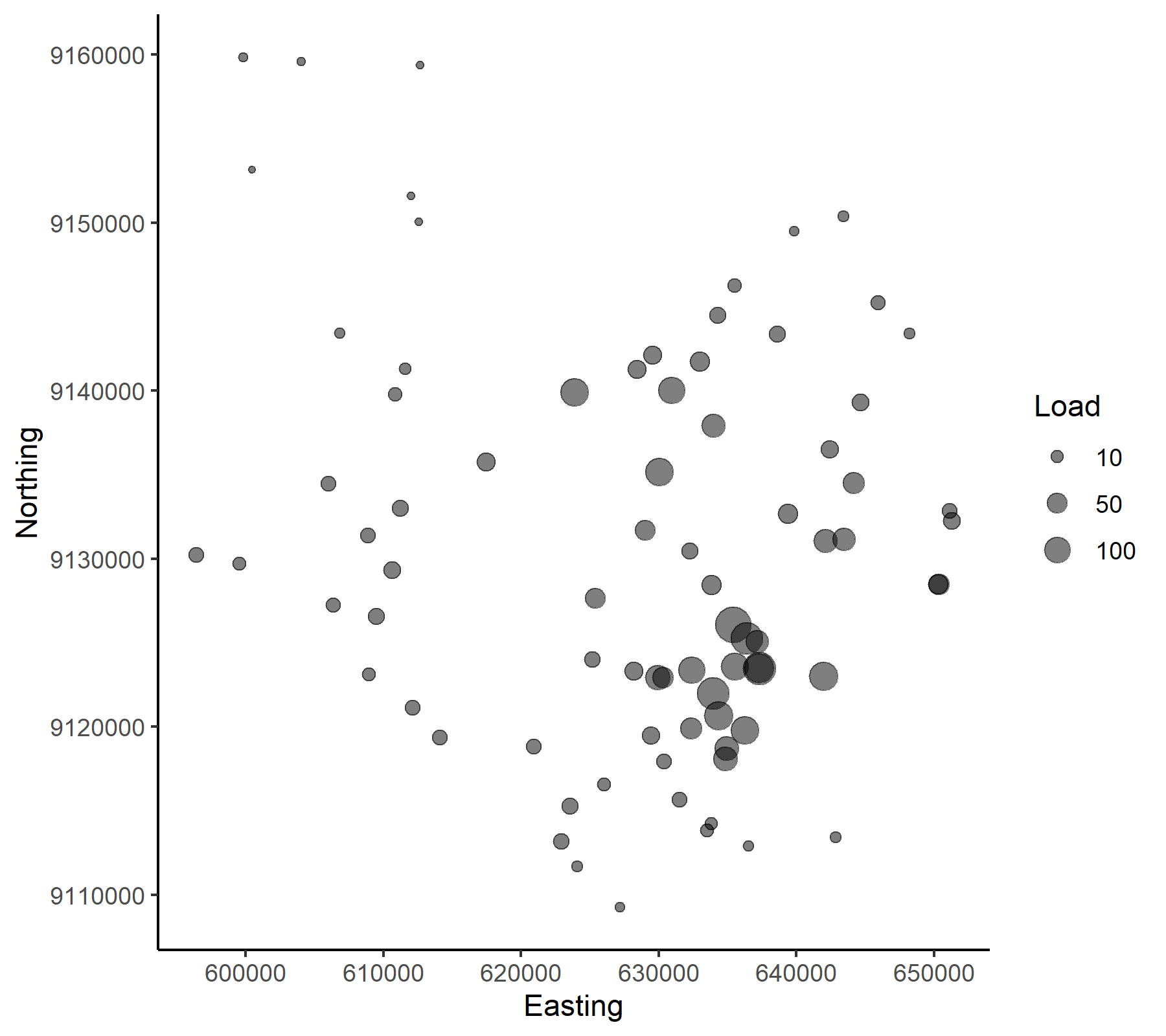}
    \end{subfigure}
        \caption{Maps of tephra load in kg/m$^2$ from two simulated datasets.}
        \label{fig:tephra_sim}
\end{figure*}

\end{document}